\documentclass[11pt, reqno]{amsart}
\usepackage{amsmath}
\usepackage{amsthm}
\usepackage{mathrsfs}
\usepackage{amssymb}
\usepackage{graphicx}

\usepackage{amsfonts}
\usepackage[mathscr]{euscript}
\usepackage[colorlinks, citecolor=blue, linkcolor=blue, urlcolor=blue]{hyperref}
\usepackage{multirow}
\usepackage{booktabs}
\usepackage{rotating}
\usepackage{subcaption}

\newcommand{\bR}{\mathbb{R}}

\newcommand{\bpsi}{\mbox{\boldmath$\psi$\unboldmath}}

\newcommand{\bbeta}{\mbox{\boldmath$\beta$\unboldmath}}

\newcommand{\bmu}{\mbox{\boldmath$\mu$\unboldmath}}
\newcommand{\blambda}{\mbox{\boldmath$\lambda$\unboldmath}}

\newcommand{\brho}{\mbox{\boldmath$\rho$\unboldmath}}

\newcommand*{\Resize}[2]{\resizebox{#1}{!}{$#2$}}%

\theoremstyle{plain}

\theoremstyle{definition}

\theoremstyle{remark}

\title[Bayesian Analysis of Censored Spatial Data ...]{Bayesian Analysis of Censored Spatial Data Based on a Non-Gaussian Model}
\author[Tadayon, V.]{Vahid Tadayon$^1$
}
\address{
 }

\begin{document}
\bibliographystyle{plainnat}
\maketitle
\footnotetext[1]{\ Department of Statistics, Eghlid Higher Education Institute, Eghid, Iran. \\
Email: vahidtadayon24@gmail.com}

\begin{abstract}
In this paper, we suggest using a skew Gaussian-log Gaussian model for the analysis of spatial censored data from a Bayesian point of view. This approach furnishes an extension of the skew log Gaussian model to accommodate to both skewness and heavy tails and also censored data. All of the characteristics mentioned are three pervasive features of spatial data.
We utilize data augmentation method and Markov chain Monte Carlo (MCMC) algorithms to do posterior calculations. The methodology is illustrated using simulated data, as well as applying it to a real data set.
\end{abstract}

\noindent 

\section{Introduction}
\label{intro}
The customary approach to spatial data modeling is to accept that the random field of interest is Gaussian. Nevertheless, this assumption always is threatened by the existence of non-Gaussianity features (such as: heavier tails or skewness) in data sets. Moreover, wrong Gaussian assumptions affect the accuracy of spatial predictions and cause bias in the resulting parameter estimates as well. So, the acceptance of the Gaussianity could be overly restrictive to interpret the quantity of interest. Therefore, there is a demand for a more flexible class of sampling models to perform efficient inference and spatial prediction in the resulting models.

For modeling only skewed spatial data without heavy tails, the most commonly adopted strategy is to use the previous model after data transformation (Tadayon and Khaledi, 2015). Even so, an appropriate transformation if there exist, may not easy to obtain (see, e.g., De Oliveira et al., 1997; Tadayon and Rasekh, 2018; Tadayon and Torabi, 2018). Kim and Mallick (2004) developed the skew Gaussian random field based on the skew normal distribution which has been introduced by Azzalini and Capitanio (1999). Zhang and El-Shaarawi (2010) recently introduced a new class of stationary process with skew normal marginal distributions. Using generalized skew Gaussian spatial field is also one of the other strategies which has been suggested by Prates et al. (2012). More details on skew spatial can be found in Rimstad and Omre (2014) and Karimi and  Mohammadzadeh (2011). 

On the other hand, the observed data may contain some isolated or grouped outliers which have extreme values compared to their neighboring observation values. Since these observations could provide additional information and knowledge about the model, they play an important role in the spatial statistics.  Measurement error is the mainspring of encountering outlying observations and belonging to a region with larger observational variance relative to the rest, is another reason. Since the non-Gaussian characteristics of the data  may be derived from these observations, in order to identify and treat with these regions, Palacios and Steel  (2006) proposed a flexible class of sampling models called Gaussian log-Gaussian (GLG). In this model which contains a scale mixed random variable, is assumed that the mixed variable is log-normally distributed. Therefore, observations with small mixed random variable belong to a region with larger observational variance. In addition, to identify individual outliers, Fonseca and Steel (2011) considered a similar mixing in the nugget effect component. However, this approach is suitable only for symmetric heavier tail distributions and will fail to handle the skewed data (Zareifard and Khaledi, 2013; Tadayon, 2018). An interesting extension, therefore, would be to develop models based on mixtures of skew normal, which would yield a highly flexible and computationally tractable parametric model that could accommodate both multimodality and extreme skewness. This extension which is compatible with Genton and Zhang (2012) remedies, is the unified skew Gaussian-log Gaussian (SUGLG) model introduced by Zareifard and Khaledi (2013). They applied the stochastic approximation expectation-maximization (SAEM) algorithm to maximize the likelihood function.

In addition, on account of some limitations of the data collection mechanism, the complete spatial data are not always available and so the quantity of interest may not be exactly observed in some sampling locations. In such setting, often the exact values can be recorded only if they fall within a specified range, which they have been called ``censored data''. One of the conventional approaches to deal with these data is to impute the censored values with ``artificial data", such as, the detection limits. However, this leads biased estimates (see, e.g., Fridley and Dixon, 2007; Tadayon, 2018). Dubrule and Kostov (1986) worked on an interpolation method which motivated by the dual formulation of kriging. They also used geostatistical approaches requiring only the second-order specification of the random field which represent a type of censored data. Utilize likelihood-based approaches for the Gaussian random field is another previous work on the analysis of spatial models in the presence of censored data which well done by Militino and Ugarte (1999). De Oliveira (2005) introduced a Bayesian approach for inference and spatial prediction based on censored data while Fridley and Dixon  (2007) incorporate the spatial correlation through an unobserved latent spatial process. Rathbun (2006) applied the Robbins-Monro (1951) stochastic approximation algorithm for estimating the parameters of a spatial regression model with left-censored observations. Eventually, Toscas (2010) proposed a modification on Bayesian approach of De Oliveira (2005) to correct the Bayesian bias in estimation and prediction of spatially correlated left-censored observations.

In all aforementioned works the inference was conducted based on the Gaussianity of the random field. However, in practice, we often observe that the exploratory data analysis shows some version of heavier tails than the normal distribution such as skewness or/and outlier region  and consequently, it violates the normality assumption. The application of some transformation is the most commonly adopted strategy. However, the existence of an appropriate transformation is one of the most important problems in this setting and the difficulty in interpretation issues is another. Tadayon and Khaledi (2015) developed and implemented a Markov chain Monte Carlo (MCMC) sampling strategy for inference and fitted the skew Gaussian model based on censored data.  In this article, we intend to fit the SUGLG model defined by Zareifard and Khaledi (2013) on a set of spatial data which could take both skewness and heavy tails simoltaneously and also contains some censored data. 

In the following section we describe the form of spatial censored data which considered in this article. The third section introduces the mixture model and derives some of its properties. Section
\ref{sec4} considers Bayesian estimation of the model parameters and describes data augmentation method as well as Bayesian prediction. Using simulated data, the identifiability of the parameters, sensitivity analysis and the application of the suggested model is studied in Section \ref{sec5}. Section
\ref{sec6} illustrates the usage of the proposed methodology on a real spatial data set which contain values of precipitation. Conclusions are presented in Section \ref{sec7}.

\section{Censored Spatial Data}
\label{sec2}
Let ${\bf y} = \left(y_1,\ldots , y_n\right)^\prime$ be a realization of ${\bf Y}=\left({ Y}\left(s_1\right),\ldots,{ Y}\left(s_n\right)\right)^\prime$, represents the data measured at the sampling locations  $s_1,s_2,\ldots,s_n$ in $R$, where $R$ is the region under study. In what follows, we assume that the mechanism that produced the censoring is uninformative, meaning that the censoring process is independent of the spatial process $Y(\cdot)$, but may be deterministic or random (De Oliveira, 2005). To argue the credibility of this assumption, consider a case in which the censoring time is independent of the survival time (Militino and Ugarte, 1999). Moreover, notice that this assumption is similar to that of data being missing at random. Thus, the observed data consist of exact observations measured at some sampling locations and interval observations measured in the rest as the result of censoring (Tadayon and Khaledi, 2015).
In this manner, to avoid some difficulties, we consider that the data of size n, consist of $m$ exact observations and $n-m$ interval observations. A reasonable methodology to define $m$ in practice is related to the nature of the data. Therefore, the data would be denoted by
\begin{eqnarray*}\label{e1}
\mathscr{D} =\left\lbrace Y\left( {\textbf{s}_{k_{j}}}\right)  = y_{k_{j}}~;~j \in J\right\rbrace \cup
\left\lbrace Y\left({\textbf{s}_{k_{i}}} \right)  \in A_{i}~;~i\in I\right\rbrace ,
\end{eqnarray*}
such that $\left\lbrace y_{k_{j}}~; j \in J\right\rbrace$ are the observed values and for any
$i\in I$, $A_i$ is the interval where $Y\left(\textbf{s}_{k_{i}}\right) $ is known to belong.
The censored interval $A_i$ can depend on the sampling situation and it varies from a location to another location.

\section{Statistical Model}
\label{sec3}
Let ${\rm Y}\left(\cdot\right)=\left\lbrace {{\rm Y}\left({{s}}\right)};s\in R \right\rbrace$ be the random field of interest where $R\subseteq \bR ^{d^*}$ and ${d^*}\geq 1$. Our starting point is the model
\begin{eqnarray}\label{e2}
  {\rm Y}\left(s\right)=\mu\left(s\right)+\frac{{\rm W}\left(s\right)}{\sqrt{\lambda\left(s\right)}}+\tau\rho
  \left(s\right)\quad,\quad\tau\ge 0,
\end{eqnarray}
with the mean surface $\mu\left(s\right)$ and the scale parameter $\tau\in\bR^+$.
$\mu\left(s\right)$ is assumed to be a linear function of
${\bf f}'\left( s\right) $, suchlike
$\mu\left(s\right)={\bf f}'\left( s\right) \bbeta$ where, ${\bf f}'\left( s\right) $ is a vector of $k$ known functions of the spatial coordinate and $\bbeta \in {\bR^k}$ is the regression coefficient vector. The term $\rho\left(s\right)$
denotes an uncorrelated Gaussian process with mean 0 and unitary variance, modeling
the so-called \textit{nugget effect}, which allows for measurement error and small-scale variation. This Further, the second-order stationary error process ${\rm W}\left(s\right)$, is a unified skew Gaussian random process corresponding to the model defined by Arellano-Valle and Azzalini (2006). On the other hand, for any vector from this random process like
${\bf W}=\left({\rm W}\left(s_1\right),\ldots ,{\rm W}\left(s_n\right)\right)^\prime$, 
we consider  that the distribution of vector $\bf W$ is
${\bf W} {\mathop=\limits^d} \alpha{\bf U}+\sigma{\bf V}$
in which,  ${\bf V}$ and ${\bf U}$ have n-variate normal distribution
${{\rm N}_n}\left({\bf 0},{C_{\theta_w}}\right)$
and truncated normal distribution ${\rm TN}_n\left({\bf 0};{\bf 0},{C_{\theta_w}}\right)$ respectively.
The $\left(i,j\right)$-th element of the $n\times n$ correlation matrix ${C_{\theta_w}}$ is ${C_{\theta_w}}\left(\|{s_i}-{s_j}\|\right)$. Thus, ${\bf W}$
has multivariate unified skew-normal (SUN) distribution\footnote{See Appendix C}
\begin{eqnarray*}\label{e3}
  {\bf W}\sim {{\rm SUN}_{n,n}}\left({\bf 0},\left({\alpha^2}+{\sigma^2}\right)
  {C_{\theta_w}},\alpha{C_{\theta_w}},0,
  {C_{\theta_w}}\right).
\end{eqnarray*}
By setting $\brho=\left(\rho\left(s_1\right),\ldots,\rho\left(s_n\right)\right)^\prime $ and $\Lambda=diag\left(\lambda_1,\ldots,\lambda_1\right) $,
equation (\ref{e2}) can be written as
\begin{eqnarray*}\label{e4}
  {\bf Y} {\mathop=\limits^d} {\rm X}\bbeta+\alpha{\Lambda^{-\frac{1}{2}}}{\bf U} +\sigma{\Lambda^{-\frac{1}{2}}}{\bf V}+\tau\brho.
\end{eqnarray*}
In addition, like Palacios and Steel (2006) we assume that $\ln{\lambda\left(s\right)}$ is a
Gaussian random field with finite-dimensional distributions:
\begin{eqnarray*}\label{e5}
     {\left({\ln\left({{\lambda_1}}\right),\ldots,\ln\left({{\lambda_n}}\right)}
    \right)^\prime}\sim {{\rm N}_n}{\left(-\frac{\nu}{2}{\rm J},\nu
    {C_{\theta_\lambda}}\right)},
\end{eqnarray*}
for ${\rm J}^\prime=\left(1,\ldots,1\right)_{n\times1}$ and $\nu>0$, which implies a lognormal distribution for each $\lambda_i$. Clearly, by tending $\nu$ to 0, $\lambda_i$ tends to 1 and so equation (\ref{e2}) reduces to unified skew Gaussian model. For simplicity, we adopt a same correlation matrices for $\textbf{W}$ and $\blambda$ but with different parameters $\theta_w$ and $\theta_\lambda$, respectively, and also we suppose that both $\textbf{W}$ and $\blambda$ have an isotropic exponential correlation function, separately. However, the first assumption might be somewhat restrictive especially in the case of spatial data without temporal replications (as identifiability problem in these parameters, if exist, can affect the estimates of parameters), but we can hope that this problem is possibly solvable when there are many observations in space. Section \ref{sec5} confirms this claim. Following Palacios and Steel (2006), we set ${\omega^2}=\frac{\tau^2}{\sigma^2}$ for simplicity and identifiability.

The model  (\ref{e2}) is able to accommodate each of individual outliers and areas of the space with an inflated variance relative to the rest, because of observations with small values $\lambda_i$ 
fall out of the mean surface.

Let $\eta^\prime=\left(\bbeta,\alpha,{\sigma^2},{\omega^2},\nu,{\theta_w},{\theta_\lambda}\right)$ and ${\bf
y}=\left(y_I,y_J\right)=\left(y_1,\ldots,y_n\right)$ be a single
realization of the considered random field with $y_i={\rm
Y}\left(s_i\right)$, then the conditional joint distribution of $\bf
y$ given $\blambda=\left(\lambda\left(s_1\right),\ldots,
\lambda\left(s_n\right)\right)^\prime$ and $\eta$ is
\begin{eqnarray*}\label{e6}
  p\left({\bf y}\left|\blambda,\eta\right.\right)\sim{{\rm SUN}_{n,n}}
  \left({\rm X}\bbeta,{\Sigma_y},\alpha
  {\Lambda^{-\frac{1}{2}}}{C_{\theta_w}},0,{C_{\theta_w}}\right).
\end{eqnarray*}
Furthermore, the likelihood function for $\eta$ given $\mathscr{D}$ is
\begin{eqnarray}\label{e7}
L\left( {\eta \left| {\mathscr{D}} \right.} \right) = \int\limits_{{{\bR^{+^n}}}} {\int\limits_{{\bR^{^n}}} {\int\limits_{{\prod\nolimits_{i \in I}} }  }}
&&p ( {{\rm{Y}}_I} \in {\Resize{0.7cm}{\prod\nolimits_{i \in I}}} {{A_i}} \left| {{y_J},{\bf{w}},\blambda ,\eta } \right. ) \nonumber\\
&&\vspace{-1cm} \times p\left( {{y_J}\left| {{\bf w},\blambda ,\eta }\right.}\right) p\left( {{\bf w}\left| {\blambda ,\eta } \right.} \right)\nonumber\\
&&\times p\left( {\blambda \left| \eta\right.}\right) d{y_I}d{\bf w} d{\blambda},
\end{eqnarray}
where ${\bf w}=\left(w\left(s_1\right),\ldots,w \left(s_n\right)\right)^\prime$,
${\Sigma_y}={\Sigma_w}+{\tau^2}{\rm{\bf I}}_n$ and ${\rm{\bf I}}_n$
is the identity matrix of dimension $n$. By setting
${C_{{JJ}_{\theta_w}}}=\left[{C_{\theta_w}}\left(\|{s_{J_i}}-{s_{J_j}}\|\right)\right]_{m\times m}$, we can easily see that
\begin{eqnarray*}\label{e8}
    {{\bf Y}{_J}}{\left|\blambda,\eta\right.}\sim  
    {{\rm SUN}_{m,m}}\left({X_J}\bbeta,{\Sigma_{y_J}},\alpha
    {{\Lambda_{J}}^{-\frac{1}{2}}}
    {C_{{JJ}_{\theta_w}}},0,{C_{{JJ}_{\theta_w}}}\right) 
\end{eqnarray*}
and the conditional distributions of interval observations
${\bf Y}{_I}\left|{y_J},\blambda,\eta\right.$ can be written as
${{\rm SUN}_{n-m,n-m}}{\prod\limits_{i \in I} {{I_{\left\{ {{y_i} \in {A_i}} \right\}}}}}$ in which
its parameters are
\begin{eqnarray*}\label{e9}
    &&%\hspace{-1cm}
    {\mu}_{y_I\left|{y_J},\blambda,\eta\right.}=X_{I}\bbeta+{\rm A}\left({{y_J}}-X_J
    \bbeta\right),   \nonumber\\
    &&%\hspace{-1cm}
    {\Sigma}_{y_I\left|{y_J},\blambda,\eta\right.}=\left({\sigma^2}+{\alpha^2}\right)
    {\Lambda_{I}^{-\frac{1}{2}}}{C_{II_{\theta_w}}}{\Lambda_{I}^{-\frac{1}{2}}}
  %  \nonumber\\&&\hspace{1cm}
    +{\tau^2}{I_{n-m}}-{\rm A}\left({\Lambda_{J}^{-\frac{1}{2}}}{C_{JI_{\theta_w}}}
    {\Lambda_{I}^{-\frac{1}{2}}}\right), \nonumber\\
    &&%\hspace{-1cm}
    {\Gamma}_{y_I\left|{y_J},\blambda,\eta\right.}={\Lambda_{I}^{-\frac{1}{2}}}{C_{IJ_{\theta_w}}}-A
    {\Lambda_{J}^{-\frac{1}{2}}}{C_{JJ_{\theta_w}}},  \nonumber\\
%    &&%\hspace{-2cm}
%    {\nu}_{y_I\left|{y_J},\blambda,\eta\right.}=      \nonumber\\
    &&%\hspace{-1cm}
    {\Delta}_{y_I\left|{y_J},\blambda,\eta\right.}={C_{JJ_{\theta_w}}}-{\alpha^2}{C_{JJ_{\theta_w}}}
    {\Lambda_{J}^{-\frac{1}{2}}}{\Sigma_{y_J}^{-1}}{\Lambda_{J}^{-\frac{1}{2}}}{C_{JJ_{\theta_w}}},\nonumber
\end{eqnarray*}
where
\begin{eqnarray}\label{e10}
    &&{\rm A}=\left({\sigma^2}+{\alpha^2}\right){\Lambda_{I}^{-\frac{1}{2}}}{C_{IJ_{\theta_w}}}
    {\Lambda_{J}^{-\frac{1}{2}}}{\Sigma_{y_J}^{-1}},   \nonumber\\
    &&{C_{IJ_{\theta_w}}}=\left[{C_{\theta_w}}\left(\|{s_{I_i}}-{s_{J_j}}\|\right)\right]_{\left(n-m\right)\times m} \nonumber
\end{eqnarray}
and the rest is $0$. It is necessary to note that $I_{\left\{{\rm A}\right\}}$ shows the indicator function of the set $\rm A$.

\section{Bayesian Inference}
\label{sec4}
To complete the Bayesian model specification, we need to select some prior distributions for all unknown parameters. One can then easily see that a convenient strategy of avoiding improper posterior distribution, is to utilize proper (but diffuse) priors. However, in this case, the insignificance of the prediction sensitivity to the hyperparameters should be established. For convenience but not necessary optimal, we assume elements of the parameter vector $\eta$ to be independent a priori, which means that
\begin{eqnarray}\label{e11}
\pi\left(\eta\right)=\pi\left(\bbeta\right) \pi\left(\alpha\right) \pi\left({\sigma^2}\right) \pi\left({\omega^2}\right) \pi\left(\nu\right) \pi\left({\theta_w}\right) \pi\left({\theta_\lambda}\right).
\end{eqnarray}
The prior distributions adopted are as follows: 
$\bbeta \sim  {{\rm N}_k}\left({\textbf{0}},{c_0}{{\rm{\textbf{I}}}_k}\right)$,
$\alpha \sim  {{\rm N}}\left({0},{c_1}\right)$,
${\sigma^{-2}}\sim  {\rm Gamma}\left({c_2},{c_3}\right)$,
${\omega^{-2}}\sim  {\rm GIG}\left(0,{c_4},{c_5}\right)$,
$\nu \sim  {\rm GIG}\left(0,{c_6},{c_7}\right)$,
${\theta_w} \sim   {\rm Exp}\left(\frac{c_8}{{ med}\left({ d}\right)}\right)$ and
${\theta_\lambda} \sim  {\rm Exp}\left(\frac{c_9}{{ med}\left({ d}\right)}\right)$,
%\begin{eqnarray*}\label{e12}
%\bbeta \sim  {{\rm N}_k}\left({\textbf{0}},{c_0}{{\rm{\textbf{I}}}_k}\right), &&
%\alpha \sim  {{\rm N}}\left({0},{c_1}\right),\qquad
%{\sigma^{-2}}\sim  {\rm Gamma}\left({c_2},{c_3}\right),\\
%{\omega^{-2}}\sim  {\rm GIG}\left(0,{c_4},{c_5}\right),&&
%\nu \sim  {\rm GIG}\left(0,{c_6},{c_7}\right),~
%{\theta_w} \sim   {\rm Exp}\left(\frac{c_8}{{ med}\left({ d}\right)}\right),~
%{\theta_\lambda} \sim  {\rm Exp}\left(\frac{c_9}{{ med}\left({ d}\right)}\right),
%\end{eqnarray*}
where the hyperparameters ${c_1},\ldots,{c_9}$ are chosen to reflect vague prior information and
${{med}\left({ d}\right)}$ is the median of all distance between the data locations. Moreover, the closed form of generalized inverse-Gaussian (GIG) distribution and some justification for utilizing the above priors can be found in  Palacios and Steel (2006).

Now, we try to derive the posterior distribution of the parameters by combining the likelihood function
(\ref{e7}) and the prior distribution  (\ref{e11}). Since a direct Bayesian analysis of the model is computationally impractical because of multiple integrals in posterior distribution, we adopt the data augmentation method and use the latent variables $\textbf{U}$ and $\blambda$ to produce some samples 
$\{({y_I^{\left(i\right)}},{{ u}^{\left(i\right)}},{\lambda^{\left(i\right)}},{\eta^{\left(i\right)}})\}_{i=1}^l$ from the extended model $p\left({y_I},{\bf U},{\blambda},\eta,{\left| \mathscr{D} \right.}\right)$ via MCMC methods. Details are given in Appendix A.

It must be noted that in many applications, the prediction of values at unsampled locations is a usual task of interest.  The spatial prediction of the response in an unobserved location $y_0$ would be based on the Bayesian predictive distribution: 
\begin{eqnarray}\label{e13}
    p\left({y_0}\left|{\bf y}\right.\right)=\int_{\prod\nolimits_{i \in I} {{A_i}} }
    {\int_{R_ + ^n} {\int_{R_ + ^p} {p\left( {{y_0}\left| {{\bf y},\bpsi ,{\bpsi _0}}
    \right.} \right)} } }
    p\left( {{\bpsi _0}\left| {{\bf y},\bpsi } \right.} \right)p\left( {\bpsi \left| {\bf y}
    \right.} \right)d{\bpsi _0}d\bpsi d{y_I},\nonumber
\end{eqnarray}
where ${\bpsi}\left(s\right)=\ln\blambda\left(s\right)$. Evidently, the full predictive distribution can not be evaluated in closed form but can be approximated using Monte Carlo samples.

\section{Simulation Studies}
\label{sec5}
The main aim of this section is to assess the performance of the proposed methodology on detecting two levels of censoring and also represent prediction values at some new unobserved locations. But since the SUGLG model may have a problem to indentify the correlation parameters (see Section \ref{sec3}), so, we first use simulation to evaluate the identifiability of these parameters.  Then we present a banal method to revalue the ability of the this model to correctly identify outlying observations as well. Throughout, results are based on every 100th draw from an MCMC chain of length 150,000 with a burn-in of 100,000 which
 provides enough output for convergence. The spatial sampling points were considered on coordinates of the data file \textit{97data.dat} available from GSLIB software (Deutsch and Journel, 1998) in which, 97 locations are taken on a pseudo-regular grid over a bidimensional region 50 by 50 miles.

\begin{figure}%[b]
 \centering
 %\hspace{-1cm}
 \includegraphics[width=0.8\textwidth, height=0.5\textwidth
					  ]
 {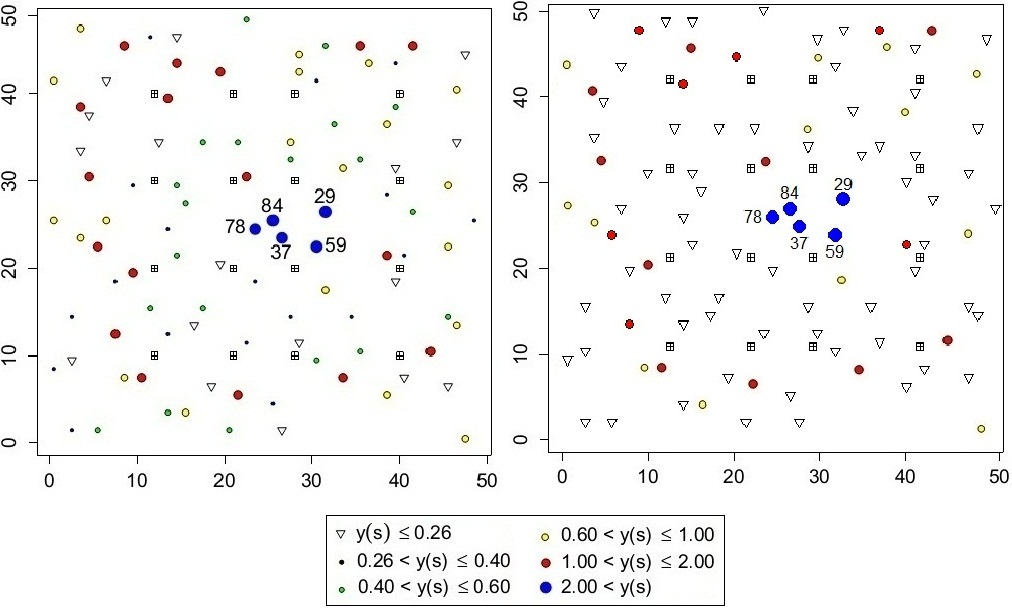}
 \caption[caption]{\footnotesize Study region based on balanced case (on the left) and extreme case (on the right) of censoring.\\\hspace{\textwidth}
 $\circ$ = exact observation.\\\hspace{\textwidth}
 $\bigtriangledown$ = left-censored observation.\\\hspace{\textwidth}
 $\boxplus$ = location which is added to spatial prediction.}
 \label{1fig}
\end{figure}

\begin{figure}%[b]
 \centering
 \includegraphics[width=6cm,height=4cm
					  ] 
					  {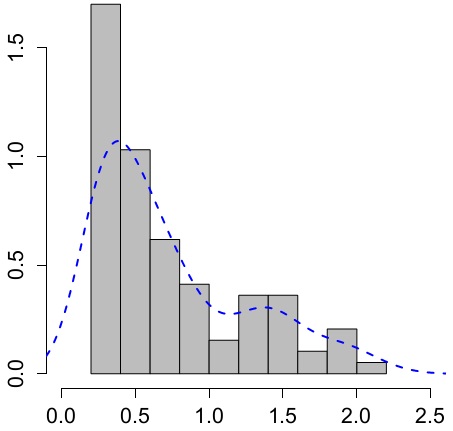}
 \caption[caption]{\footnotesize Histogram with the nonparametric density estimator of the generated data.}
 \label{2fig}
\end{figure}

First of all, we appoint a lattice locations
\begin{eqnarray}\label{lat1}
\left\lbrace{12,21,28,40}\right\rbrace\times\left\lbrace{10,20,30,40}\right\rbrace,
\end{eqnarray}
 as a hold-out data set to evaluate the performance of the model in spatial prediction and then data are generated on these 113 locations based on the sampling model
\begin{eqnarray*}
	{\bf Y}={\beta_0}+\alpha{\bf U}+\sigma{\bf V}+\tau\rho,
\end{eqnarray*}
with a constant mean surface $\beta_0=0$, $\alpha=3, \sigma=1, \tau=0.1$ and ${\theta_\lambda}={\theta_w}=0.5$. After that, a cluster of outliers at locations 29, 37, 59, 78 and 84 has been created by imposing two units to the simulated values. Then, by leaving the simulated value in
the lattice locations (\ref{lat1}), the model will be fitted on the rest (of size 97).
For model validation, we now implement two levels of left-censored design, one 17.5$\%$ and other 67$\%$ as a \textit{balanced} and \textit{extreme} cases of censoring respectively, such that in the first of them, the 17th ordered observation will be substituted for 1st up to 17th smallest values and in the second, 1st up to 65th smallest observations will be replaced with 65th ordered observation. Fig. \ref{1fig} shows a schematic description of the region that displays the sampling
locations as well as the region with inflated variance and censored locations. As an exploratory analysis, Fig. \ref{2fig} which shows the histogram with the nonparametric density estimator of generated data, confirm the existence of an outlier region in these data as well as skewness.

Since the inference may be challenging in identifying the correlation parameters $\theta_w$ and $\theta_\lambda$, here, we focus on this problem to see what extent information about these parameters can be recovered from data. This study will be done just based on the balanced case of censoring because we do not expect a high ability to identify the parameters in the extreme case. To check the aforementioned problem, three data sets were generated with $\theta_w=0.25, 1, 1.5$ and $\theta_\lambda=0.1, 0.7, 1.1$ (in which, the other values have been fixed) and the estimate of the model parameters are obtained. However there are quite moderate sample size, the results which are reported in Table \ref{1tab}, indicates that there is no significant problem on the identifying of the correlation parameters and so the data almost allow for acceptable inference on these parameters.

At the moment, we concentrate on the performance of the Bayesian  estimation, in which we must do one of these actions: choose some benchmark values for the various hyperparameters used in the priors, using empirical Bayes or employing hierarchical Bayes. To facilitate the computations, we consider the first method. But, we must investigatethe robustness of the posterior results under changes in the prior hyperparameters. It must be noted that we do not expect stability of the posteriors per se, but rather our interest is often in prediction. For this purpose, we employ the \textit{relative change} criterion, which is the absolute value of the induced change in the marginal posterior mean of each parameter divided by the standard deviation computed under the benchmark prior. Table \ref{2tab} which lists the alternative prior hyperparameters as well as the maximum relative change recorded for each parameter, shows no significant changes.

\begin{table}%[b]
\centering
\caption{\footnotesize Identifiability of the correlation parameters based on the balanced case of censoring.}
\begin{tabular}{l l c c c c c}
\toprule
% g & g & g  \\ \hline
\multirow{2}{*}{$\theta_w$}
 & True value & 0.2 & 1 & 1.5 \\
 & Estimated value & 0.98 & 1.15 & 1.41  \\
\hline
\multirow{2}{*}{$\theta_\lambda$} 
 & True value & 0.1 & 0.7 & 1.1  \\ 
 & Estimated value & 0.14 & 0.81 & 0.98 \\
\bottomrule
\end{tabular}
\label{1tab}
\end{table} 

\begin{table}%[b]
\centering 
\caption{\footnotesize Prior Sensitivity Analysis: Setup of the Experiment and Maximum Relative Change (MRC). }
%Prior Sensitivity Analysis: benchmark and alternative values for the hyperparameter and maximum relative change (MRC) in the marginal posterior of each parameter.}
\begin{tabular}{c c c c c c c c c c} 
\toprule 
%\multirow{3}{*}{\begin{sideways}Arg.\end{sideways}}
	& & & &\multicolumn{2}{c}{\footnotesize{MRC}} \\
\cmidrule(rl){5-6} 
			& \footnotesize{Hyperparameter}
			& \footnotesize{Benchmark}
			& \footnotesize{Alternative values}
			&\footnotesize{$17.5\%$}
			&\footnotesize{$67\%$}\\ 
\cmidrule{1-6} 
${\beta_0}$		& $c_0$ & $10^4$ & ${10^2}$ , ${10^6}$ & 0.0611	& 0.0926
\\ 
$\alpha$           & $c_1$ & $10^5$ & ${10^3}$ , ${10^7}$ & 0.132		& 0.1915		
\\

${\sigma^2}$  	& $\left({c_2},{c_3}\right) $ & $\left({10^{-6}},{10^{-6}}\right)$  
		& $\left({10^{-4}},{10^{-8}}\right)$,$\left({10^{-8}},{10^{-4}}\right)$
		& 0.1709	& 0.1135
\\
${\omega^2}$	& $\left({c_4},{c_5}\right) $ & $\left(0.1,9 \right)$
		& $\left(0.7,1.5 \right)$ , $\left(0.5,0.7 \right)$
		& 0.3147	& 0.652	
\\
$\theta_w$		& $c_8$ & $0.7$ & $0.4$ , $1.4$ 		& 0.393 	& 	0.4117
\\
$\theta_\lambda$	& $c_9$ & $0.7$ & $0.35$ , $1.60$	& 0.421 	& 0.4705	
\\
$\nu$				& $\left({c_6},{c_7}\right) $ &  $\left(0.5,1.5 \right)$
		& $\left(0.30,1.0\right)$ , $\left(0.75,0.8 \right)$	& 0.8003		&  0.96		
\\
%\midrule 
\bottomrule
\end{tabular}
%\captionsetup{justification=centering}
\label{2tab}
\end{table}

Therefore, the sensitivity of posterior can be partitioned into three groups: First, the prior on
$\left( {\beta_0},\alpha,{\sigma^2}\right) $ which does not play a vital role, $\left({\omega^2},{\theta_w},{\theta_\lambda}\right)$ where the data is reasonably informative on these parameters as the second and the last is $\nu$ by the largest influence of prior changes. Since there is relatively little direct information in the data on $\nu$, choice of prior is quite important. However, the adoption of hierarchical or empirical Bayes can be helpful.
%To conquest this promlem, we must adopt an approach to decrease the sensitivity of $\nu$ marginal posterior. Table \ref{30tab} shows the maximum relative changes of marginal posteriors based on the benchmark hyperparameters which are listed in Table \ref{2tab} but, by adoption of hierarchical Bayes method.
%\begin{table}[b]
%\centering
%\begin{tabular}{c c c c c c c c c }
%\toprule
% 
%&\multirow{1}{*}{Parameter}&${\beta_0}$&$\alpha$&${\sigma^2}$&${\omega^2}$&$\theta_w$&
%$\theta_\lambda$&$\nu$ \\
%%\hline
%\cmidrule(lr){1-9} 
%\multirow{3}{*}{\begin{sideways}MRC\end{sideways}}
% & $17.5\%$  & 0.0396 & 0.1312 & 0.167 & 0.2451 & 0.38 & 0.4 & 0.5279  \\ 
%	\\
% & $67\%$  & 0.088 & 0.179 & 0.105 & 0.5801 & 0.3903 & 0.3621 & 0.772 \\
%\bottomrule
%\end{tabular}
%\caption{Prior Sensitivity Analysis: benchmark and alternative values for the hyperparameter and maximum relative change (MRC) in the marginal posterior of each parameter.}
%\label{30tab}
%\end{table}

Table \ref{3tab} summarizes  the results of  fitting  four models:  
\begin{itemize}
\item Gaussian (GAUS):
${\rm Y}\left(s\right)={\beta_0}+\tau\rho\left(s\right)$,
\item Unified Skew Gaussian (SUG):
${\rm Y}\left(s\right)={\beta_0}+{\rm W}\left(s\right)+\tau\rho \left(s\right)$,
\item Gaussian-Log Gaussian (GLG):
${\rm Y}\left(s\right)={\beta_0}+{{\lambda\left(s\right)}}^{-\frac{1}{2}}+\tau\rho \left(s\right)$,
\item Unified Skew Gaussian-Log Gaussian (SUGLG):\vspace{-0.2cm}
\begin{eqnarray*}
{\rm Y}\left(s\right)={\beta_0}+\frac{{\rm W}\left(s\right)}{\sqrt{\lambda\left(s\right)}}+\tau\rho
  \left(s\right).
\end{eqnarray*}
\end{itemize}%\vspace{-0.3cm}
According to this table, it is found that the SUGLG model has a better performance. Moreover, to compare four models in prediction, we compute the root of mean-square error of the predicted response values on the determined lattice locations. Table \ref{4tab} suggets the desirability of using the SUGLG model.
\begin{table}%[b]
\centering
\caption{\footnotesize The estimation of the parameters in four models: GAUS, 
SUG, GLG and SUGLG.}
\begin{tabular}{c c c c c c c c c c c c c}
\toprule
&&&${\beta_0}$&$\alpha$&${\sigma^2}$&${\omega^2}$&$\theta_w$&
$\theta_\lambda$&$\nu$  \\
\cmidrule(lr){4-10} 
&&\multirow{1}{*}{True value}
		& 0 & 3 & 1 & 0.1 & 0.5 & 0.5 & 1\\ 
\cmidrule(lr){1-10} 
\multirow{8}{*}{\begin{sideways}{ Estimated values} \end{sideways}}&
\multirow{2}{*}{\begin{sideways}{\scriptsize GAUS} \end{sideways}}
& $17.5\%$ & 1.67 & - & 2.17 & 0.9 & 0.38 & 0.15 & - \\ 
&& $67\%$ & 2.03 & - & 1.86 & 0.91 & 0.46 & 0.4 & - \\ 
\cmidrule(lr){2-10} 
&
\multirow{2}{*}{\begin{sideways}{\scriptsize SUG} \end{sideways}}
& $17.5\%$ & 1.21 & 2.03 & 1.68 & 0.83 & 0.45 & 0.12 & - \\ 
&& $67\%$ & 1.42 & 1.35 & 1.4 & 0.71 & 0.58 & 0.39 & - \\ 
\cmidrule(lr){2-10} 
&
\multirow{2}{*}{\begin{sideways}{\scriptsize GLG} \end{sideways}}
& $17.5\%$ & 0.98 & - & 1.74 & 0.66 & 0.22 & 0.34 & 0.51 \\ 
&& $67\%$ & 1.5 & - & 1.38 & 0.59 & 0.55 & 0.59 & 1.23\\ 
\cmidrule(lr){2-10} 
&
\multirow{2}{*}{\begin{sideways}{\scriptsize SUGLG} \end{sideways}}
& $17.5\%$ & 0.05 & 2.77 &1.33 & 0.2 & 0.56 & 0.45 & 1.12 \\ 
&& $67\%$ & 0.8 & 2.13 & 1.15 & 0.23 & 0.46 & 0.53 & 0.92 \\ 
\bottomrule
\end{tabular}
\label{3tab}
\end{table}

\begin{table}%[b]
\centering
\caption{\footnotesize The root of mean square error (RMSE) of the predicted response values based on GAUS, SUG, GLG and SUGLG models.}
\begin{tabular}{c c c c c c c}
\toprule
 & & \footnotesize GAUS & \footnotesize SUG & \footnotesize GLG & \footnotesize SUGLG \\
\hline
\multirow{2}{*}{\begin{sideways}{\scriptsize RMSE} \end{sideways}}
 &  $17.5\%$& 0.191 & 0.083 &0.095& 0.031  \\ 
 & $67\%$  & 0.211 & 0.187 & 0.166& 0.102 \\
\bottomrule
\end{tabular}
\label{4tab}
\end{table}

\begin{figure}%[b]
 \centering
 \includegraphics[width=0.7\textwidth, height=0.2\textwidth
					  ]
 {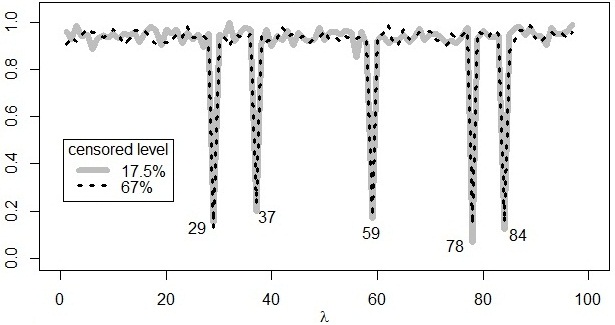}
 \caption{\footnotesize  Mean of the posterior values $\lambda_i$'s for $i=1,\ldots,97$. Small values of $\lambda_i$'s
 lead to large prediction in response.}
 \label{3fig}
\end{figure}

Finally, by plotting the Bayesian estimate of $\lambda_i$'s for $i=1,\ldots,n$, it can be seen that
how posterior values of $\lambda_i$'s might be used to identify outliers (because of the effect of small values of $\lambda_i$'s in the denominator, see equation  (\ref{e2})).
Fig. \ref{3fig} which shows mean of the posterior values $\lambda_i$'s for $i=1,\ldots,97$, confirms the existence of an outlier region in the clustered locations 29, 37, 59, 78 and 84.

\section{Rainfall Data Analysis }
\label{sec6}
\begin{figure}%[b]
 \centering
 \includegraphics[width=0.9\textwidth, height=0.4\textwidth
					  ] 
					  {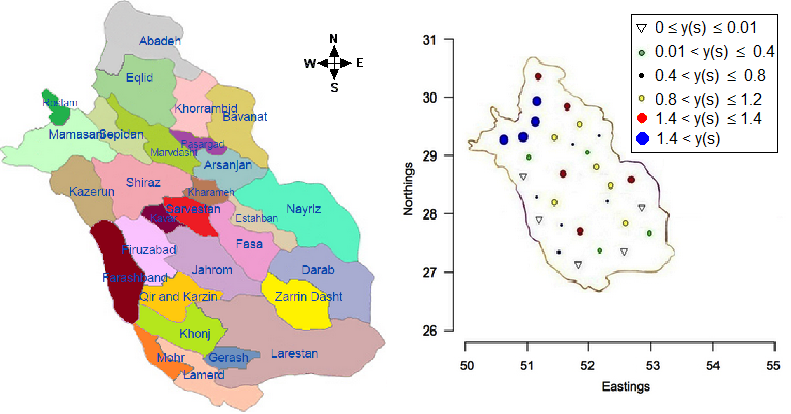}
 \caption{\footnotesize Position of the 30 stations and their respective rainfall amounts.}
 \label{4fig}
\end{figure}

The analyzed data set is comprised of rainfall amounts, measured in inch and accumulated on the first and coldest Wednesday in December 2012 at 30 stations located in Fars province of Iran. A schematic description of the region, the stations and rainfall amounts is shown in Fig. \ref{4fig}. In this figure it can be seen that there is a region with larger observation relative to the rest, along the north-eastern of the study region. In this data set (which are presented in Appendix B), the zero values are reported for
five stations. Nevertheless, since the nonzero values are reported for the countrysides of these stations by residents and the stations operators, we consider their values as censored with censored intervals $\left[0,0.01\right)$, where the value 0.01 is the least value of precipitation in their countrysides. These locations are indicated by "$\bigtriangledown$" as a left-censored observations in Fig. \ref{4fig}.

\begin{figure}%[b]
        \centering
        \begin{subfigure}[t]{0.5\textwidth}
                \centering
                \includegraphics[width=0.6\textwidth, height=0.6\textwidth]{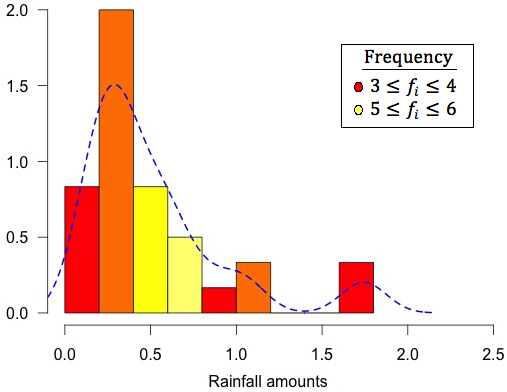}
                \caption{}
                \label{fig:gull}
        \end{subfigure}%
        ~ %add desired spacing between images, e. g. ~, \quad, \qquad etc. 
          %(or a blank line to force the subfigure onto a new line)
        \begin{subfigure}[t]{0.45\textwidth}
                \centering
                \includegraphics[width=0.6\textwidth, height=0.6\textwidth]{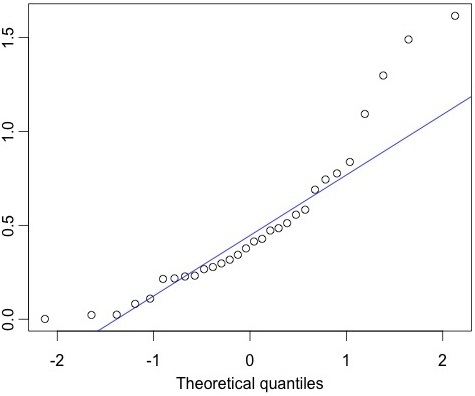}
                \caption{}
                \label{fig:tiger}
        \end{subfigure}\\ \vspace{-.1cm}
        ~ %add desired spacing between images, e. g. ~, \quad, \qquad etc. 
          %(or a blank line to force the subfigure onto a new line)
        \begin{subfigure}[t]{0.45\textwidth}
                \centering
                \includegraphics[width=0.67\textwidth, height=0.67\textwidth]{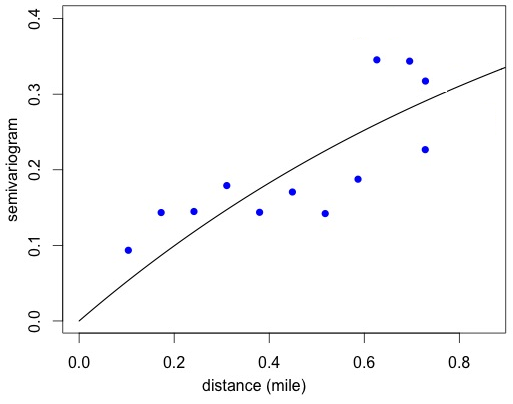}
                \caption{}
                \label{fig:mouse1}
        \end{subfigure}
        \begin{subfigure}[t]{0.5\textwidth}
                \centering
                \includegraphics[width=0.6\textwidth, height=0.6\textwidth]{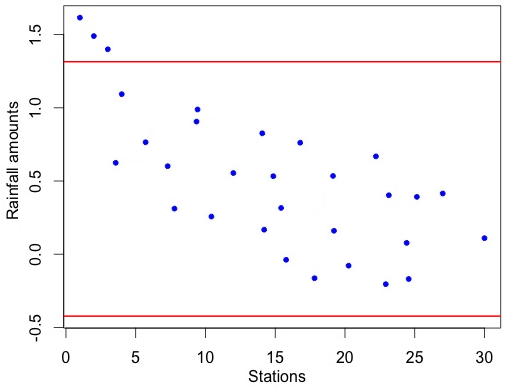}
                \caption{}
                \label{fig:mouse2}
        \end{subfigure}
        \caption{\footnotesize{Exploratory data analysis (EDA) of rainfall dataset:
        (a) Histogram with the nonparametric density estimator of generated data,
        (b) Q-Q plot,
        (c) Empirical Semivariogram,
        (d) The Haining limitations.}}\label{4in1}
\end{figure}

First of all, we performed the exploratory data analysis (EDA) such as the histogram, Q-Q plot, 
the Shapiro-Wilk’s test, the empirical semivariogram and the Haining (1991) method  to detect outlier observations. Some of those results are presented in Fig. \ref{4in1}.
The histogram and Q-Q plot suggest that the data set has moderately right-skewed distribution which was confirmed by using the Shapiro-Wilk test where reported in Table \ref{5tab}. In addition, comparing the Haining bounds and the response values, indicate a region in the northings of studying space with larger observational variance relative to the rest, where we observed large amounts of precipitation. Finally, it is clear from the empirical semivariograms that there exists a strong spatial correlation as well as a nugget effect in the data set.

\begin{table}%[b]
\centering 
\caption{\footnotesize Shapiro-Wilk's normality test.}
\begin{tabular}{c c} 
\toprule
Shapiro-Wilk & P-value\\ 
\cmidrule(lr){1-2}
$0.8765$ & $0.002343$\\ 
\bottomrule
\end{tabular}
\label{5tab}
\end{table}

Before starting to compute estimations and spatial predictions, we mention that, since
explanatory analysis of the data did not show any significant relation between response value
and the spatial locations, the mean function is assumed to be constant, so we have just $\beta_0$ and
thus $k = 1$. This exploratory data analysis suggests that SUGLG model is a suitable option for doing the analysis. To perform Bayesian inference, benchmark values of the hyperparameter have chosen  same as the given values in Table \ref{2tab}. The results is represented in 
Table \ref{6tab}. Although the DIC and the LMPL ($= \sum\nolimits_{i = 1}^n {\log \left( {CP{O_i}} \right)} $)\footnote{The conditional predictive ordinate (CPO): The CPO model comparison is a Bayesian cross-validation approach (e.g., Geisser, 1993)} criterion present a versus results in comparison between two SUG and GLG models in some situations, but both of them confirm the better performance of the SUGLG model.

\begin{figure}%[b]
        \centering
        \begin{subfigure}[t]{0.18\textwidth}
                \centering
                \includegraphics[width=\textwidth]{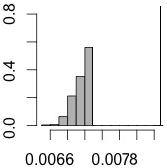}
                \caption{}
                \label{a}
        \end{subfigure}%
        ~ %add desired spacing between images, e. g. ~, \quad, \qquad etc. 
          %(or a blank line to force the subfigure onto a new line)
        \begin{subfigure}[t]{0.18\textwidth}
                \centering
                \includegraphics[width=\textwidth]{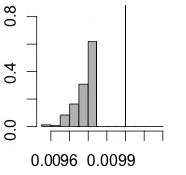}
                \caption{}
                \label{b}
        \end{subfigure}
        ~ %add desired spacing between images, e. g. ~, \quad, \qquad etc. 
          %(or a blank line to force the subfigure onto a new line)
        \begin{subfigure}[t]{0.18\textwidth}
                \centering
                \includegraphics[width=\textwidth]{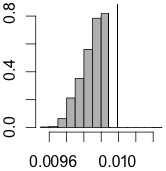}
                \caption{}
                \label{c}
        \end{subfigure}
        \begin{subfigure}[t]{0.18\textwidth}
                \centering
                \includegraphics[width=\textwidth]{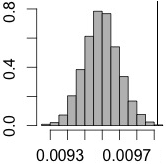}
                \caption{}
                \label{d}
        \end{subfigure}
        \begin{subfigure}[t]{0.18\textwidth}
                \centering
                \includegraphics[width=\textwidth]{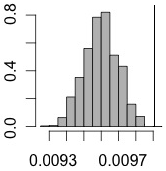}
                \caption{}
                \label{e}
        \end{subfigure}
        \caption{\footnotesize Estimates of the posterior distribution of ${\rm Y}\left({s_i}\right)$
        for five censored observations. Vertical lines are placed at the respective censoring limits. (a) to (e) correspond to  Darab, Firuzabad, Lamerd, Larestan and Qir-o-Karzin, respectively.}
        \label{5in1}
\end{figure}

Morover, the prediction of rainfall amounts in the five mentioned stations, is presented in Table
\ref{7tab}.
One of the attractive feature of the Bayesian approach of course, is that it allows to make
inference about the censored values, and in particular to quantify uncertainty about them. Fig. \ref{5in1} displays the histograms of the posterior samples of the censored values, where vertical lines are placed at their respective censoring limits. It is clear that for some of these locations there is a lot of uncertainty
about the true value, while for others the respective censoring limit is a poor estimate of the true value.

\begin{figure}%[b]
 \centering
 \includegraphics[width=\textwidth, height=0.4\textwidth]{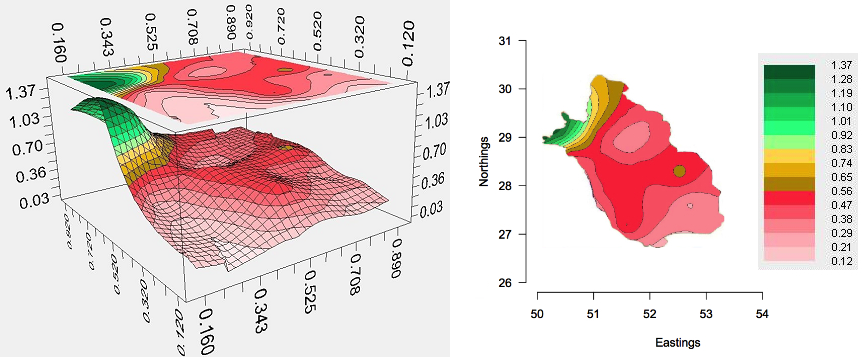}
 \caption{\footnotesize The contour map corresponding to the predictive mean for rainfall dataset.}
 \label{9fig}
\end{figure}

\begin{figure}%[b]
 \centering
 \includegraphics[width=0.9\textwidth, height=0.5\textwidth]{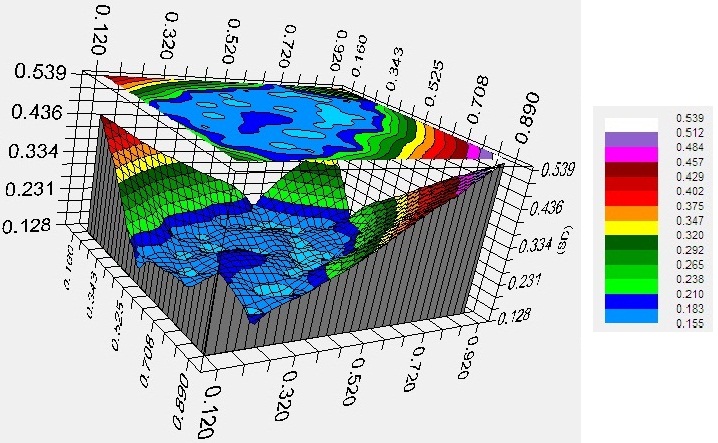}
 \caption{\footnotesize Topographic data: standard deviation surface of predictive mean.}
 \label{9p13}
\end{figure}

Finally, the contour map corresponding to the predictive mean surface under the SUGLG case, is shown
in Fig. \ref{9fig} which are computed from the predictive distribution over a regular grid of $20 \times 20$. According to this figure, the predictions are highest in the area of inflated variance that contains the aforementioned observations. Fig. \ref{9p13} presents the standard deviation surfaces for the SUGLG case as well.

\begin{table}%[b]
\centering
\caption{\footnotesize Estimation (Est) and Standard deviation (SD)
of the parameters in four models: GAUS, SUG, GLG and SUGLG with the DIC and the LPML criterion to comparison these models.}
\begin{tabular}{c c c c c c c c c c c}
\toprule
& &${\beta_0}$&$\alpha$&${\sigma^2}$&${\omega^2}$&$\theta_w$&$\theta_\lambda$&$\nu$  & {DIC} & {LPML}  \\ 
\cmidrule(lr){1-11}
\multirow{2}{*}{\begin{sideways}{\scriptsize GAUS} \end{sideways}}
&Est& 0.006 & -     & 2.13 & 0.291   & 0.036 & 0.521 & -      & \multirow{3}{*}{57.3}&\multirow{3}{*}{19.4} \\  
&SD & 1.71 & -     & 1.03 & 0.84   & 2.17 & 1.97 & -       \\  
\cmidrule(lr){1-11} 
\multirow{2}{*}{\begin{sideways}{\scriptsize SUG} \end{sideways}}
&Est& 0.043 & 2.76 & 1.98 & 0.284 & 0.251 & 0.56 & -      & \multirow{3}{*}{40.7} & \multirow{3}{*}{31.1} \\  
&SD& 1.65 &1.19   & 1.11 & 0.761   & 2.061 & 2.01 & -       \\  
\cmidrule(lr){1-11} 
\multirow{2}{*}{\begin{sideways}{\scriptsize GLG} \end{sideways}}
&Est& 0.044 & -     & 1.9 & 0.276  & 0.232 & 0.473 & 0.32 & \multirow{3}{*}{39.6} & \multirow{3}{*}{30} \\ 
&SD& 1.13 & -     & 1.86 & 0.874   & 1.81 & 1.68 & 0.24    \\  
\cmidrule(lr){1-11} 
\multirow{2}{*}{\begin{sideways}{\scriptsize SUGLG} \end{sideways}}
&Est& 0.011 & 2.22 & 1.4 & 0.197  & 0.461 & 0.465 & 2.61 & \multirow{3}{*}{11.4} & \multirow{3}{*}{43.7}\\  
&SD& 0.33 &0.95    & 0.41 & 0.29   & 0.925 & 0.71 & 0.13      \\  
\bottomrule
\end{tabular}
\label{6tab}
\end{table}

\begin{table}%[b]
\centering
\caption{\footnotesize Bayesian spatial prediction of rainfall values in five stations.}
\begin{tabular}{c c c}
\toprule
& Observed value & Predicted value \\
\cmidrule(lr){1-3}
Darab &0& 0.0070121\\
Firuzabad &0& 0.009761 \\
Lamerd&0&0.0099013\\
Larestan&0&0.0095678\\
Qir-o-Karzin&0&0.0095411\\
\bottomrule
\end{tabular}
\label{7tab}
\end{table}

\section{Conclusions}
\label{sec7}
In this article we have considered a unified skew Gaussian-log Gaussian model to analyze of non-Gaussian random fields based on censored data. Overall, the Bayesian framework,  data
augmentation method and MCMC algorithms have been extended for parameter estimation.
However,  using this algorithm entail more complication in the computations, recently, some studies have been performed to find replacing framework and it has led to  some new methods such as Variational Bayes’ method (Ren et al. 2011). Although this method requires more complex theoretic calculations, it could increase the speed of calculations. So, assessing the performance of this method in Bayesian inference of Non-Gaussian random fields in the presence of censored values is an interesting area to investigate in further research.

%%%%%%%%%%%%%%%%%%%%%%%%%%%

%%%%%%%%%%%%%%%%%%%%%5

\appendix{}
\section*{Appendix A: The Full Conditional Distributions}
\label{AppA}
In sequel, we try to present the full conditional distributions of all unknown quantities
$p\left({y_I},{\bf U},{\blambda},\eta,{\left| \mathscr{D}\right.}\right)$ in the Gibbs sampler framework. Moreover, to facilitate the computations we put
the notation ${\eta_{-\phi}}$ to show the vector $\eta$ without the element $\phi$ and set $Z={\bf y}-X\bbeta$.

\begin{itemize}
\item[$\bullet$] \textbf{Full conditional distribution of censored values}\\
By considering the partition of ${\bf u}=\left(u_I,u_J\right)$ we have
\begin{eqnarray*}
{\bf Y}{_I}\left|{\bf u},{v},\blambda,\eta,\mathscr{D}\right.\sim
{{\rm N}_{n-m}}\left({\mu^*},{\Sigma^*}\right){\prod
\limits_{i \in I} {{I_{\left\{ {{y_i} \in {A_i}} \right\}}}}},
\end{eqnarray*}
where,
\begin{eqnarray*}
{\mu^*}&=&{X_I}\bbeta+\alpha{\Lambda_{I}^{-\frac{1}{2}}}{{u}_I}
+ {\Sigma_{{y_I}{y_J}}}{\Sigma_{{y_J}{y_J}}^{-1}}
\left({y_J}-{X_J}\bbeta-\alpha{\Lambda_{J}^{-\frac{1}{2}}}{u_J}\right),\\
{\Sigma^*}&=&{\Sigma_{{y_I}{y_I}}}-{\Sigma_{{y_I}{y_J}}}{\Sigma_{{y_J}{y_J}}^{-1}}
{\Sigma_{{y_J}{y_I}}}.
\end{eqnarray*}
which is a $(n-m)$-variate truncated normal distribution and
so it can easily generated by standard algorithm.
%============================================
\item[$\bullet$] \textbf{Full conditional distribution of latent variable ${\bf U}$}
\begin{eqnarray*}
p\left( {{\bf U}\left| {{\bf y},\blambda ,\eta } \right.} \right) &\propto & p\left( {{\bf y}\left| {{\bf u},\blambda ,\eta } \right.} \right)p\left( {{\bf U}\left| {\blambda ,\eta } \right.} \right)\pi\left( {\blambda \left| \eta \right.} \right)\\
&\propto & \exp \left\{ { - \frac{1}{2}{{\left( { Z - \alpha {\Lambda ^{ - \frac{1}{2}}}{\bf U}} \right)}^\prime }\Sigma _y^{ - 1}\left( {Z - \alpha {\Lambda ^{ - \frac{1}{2}}}{\bf U}} \right)} \right\}\\
&\times &\exp \left\{ { - \frac{1}{2}{\bf U}'C_{\theta_w} ^{ - 1}{\bf U}} \right\}\\
&\propto &\exp \left\{ { - \frac{1}{2}\left( {{\bf U}'\left( {C_{\theta_w} ^{ - 1} + {\alpha ^2}{\Lambda ^{ - \frac{1}{2}}}\Sigma _y^{ - 1}{\Lambda ^{ - \frac{1}{2}}}} \right){\bf U} + 2\alpha {\bf U}'{\Lambda ^{ - \frac{1}{2}}}\Sigma _y^{ - 1} {Z } } \right)} \right\},
\end{eqnarray*}
which means that ${\bf U}\left|{{\bf y},\blambda,\eta}\right.\sim {{\rm TN}_{n}}
\left({\bf 0};{\mu_{u}^*},{H_u^{-1}}\right)$ by 
\begin{eqnarray*}
{\mu_{u}^*}&=&{H_u^{-1}}\left({\Lambda^{-\frac{1}{2}}}{\Sigma_{y}^{-1}}\right)Z,\\
{H_u}&=&\frac{1}{\alpha^2}{C_{\theta_w}^{-1}}+{\Lambda^{-\frac{1}{2}}}{\Sigma_{y}^{-1}}
{\Lambda^{-\frac{1}{2}}}.
\end{eqnarray*}
%============================================
\item[$\bullet$] \textbf{Full conditional distribution of regression coefficients}
\begin{eqnarray*}
\pi\left( {\bbeta \left| {{y_I},{\bf u},\blambda ,{\eta _{ - \bbeta }},\mathscr{D}} \right.} \right) &\propto & \pi\left( {\bbeta \left| {{\bf y},{\bf u},\blambda ,{\eta _{ - \bbeta }}} \right.} \right)\\
&\propto & p\left( {{\bf y}\left| {{\bf u},\blambda ,\eta } \right.} \right)\pi\left( \bbeta \right)\\
&\propto &\exp \left\{ { - \frac{1}{2}\left[ {{\bf y}'\Sigma _y^{ - 1}X\bbeta - \bbeta X'\Sigma _y^{ - 1}{\bf y} + \bbeta 'X'\Sigma _y^{ - 1}X\bbeta } \right.} \right. \\
&&\hspace{0.73cm}+\bbeta 'X'\Sigma _y^{ - 1}\alpha {\Lambda ^{ - \frac{1}{2}}}{\bf u} + \alpha {\bf u}'{\Lambda ^{ - \frac{1}{2}}}\Sigma _y^{ - 1}X\bbeta \\
&&\hspace{0.73cm} \left. {+\bbeta '{{\left( {{c_0}{I_k}} \right)}^{ - 1}}\left. \bbeta \right]} \right\}\\
&\propto &\exp \left\{ { - \frac{1}{2}\left[ {\bbeta '\left( {X'\Sigma _y^{ - 1}X + {{\left( {{c_0}{I_k}} \right)}^{ - 1}}} \right)\bbeta } \right.} \right.\\
&&\hspace{0.73cm}\left. {\left. { - 2\bbeta 'X'\Sigma _y^{ - 1}\left( {{\bf y} - \alpha {\Lambda ^{ - \frac{1}{2}}}{\bf u}} \right)} \right]} \right\}.
\end{eqnarray*}
So, $\bbeta\left|{\bf y},{\bf u},{\blambda},{\eta_{-\bbeta}}\right.\sim{{\rm N}_k}\left({\mu_{\bbeta}^*},{{\bf H}_{\bbeta}^{-1}}\right) $ with
\begin{eqnarray*}
&&{\mu_{\bbeta}^*}={{\bf H}_{\bbeta}^{-1}}{X^\prime}{\Sigma_{y}^{-1}}\left({\bf y}-\alpha
{\Lambda^{-\frac{1}{2}}}{\bf u}\right),\\
&&{{\bf H}_{\bbeta}}={X^\prime}{\Sigma_{y}^{-1}}+\frac{1}{c_0}{{\rm I}_k}.
\end{eqnarray*}
%============================================
\item[$\bullet$] \textbf{Full conditional distribution of parameter $\alpha$}
\begin{eqnarray*}
\pi\left( {\alpha \left| {{y_I},{\bf u},\blambda ,{\eta _{ - \alpha }},\mathscr{D}} \right.} \right) &\propto &\pi \left( {\alpha \left| {{\bf y},{\bf u},\lambda ,{\eta _{ - \alpha }}} \right.} \right)\\
&\propto &p\left( {{\bf y}\left| {{\bf u},\blambda ,\eta } \right.} \right)\pi \left( \alpha \right)\\
&\propto &\exp \left\{ { - \frac{1}{2}\left[ { -{\bf y}'\Sigma _y^{ - 1}\alpha {\Lambda ^{ - \frac{1}{2}}}{\bf u} + \bbeta 'X'\Sigma _y^{ - 1}\alpha {\Lambda ^{ - \frac{1}{2}}}{\bf u}} \right.} \right.\\
&&\hspace{0.73cm} -{\bf u}'{\Lambda ^{ - \frac{1}{2}}}\alpha \Sigma _y^{ - 1}{\bf y} + \alpha {\bf u}'{\Lambda ^{ - \frac{1}{2}}}\Sigma _y^{ - 1}X\bbeta \\
&&\hspace{0.73cm}\left. {\left. { + {\alpha ^2}{\bf u}'{\Lambda ^{ - \frac{1}{2}}}\Sigma _y^{ - 1}{\Lambda ^{ - \frac{1}{2}}}{\bf u} + \frac{1}{{{c_1}}}{\alpha ^2}} \right]} \right\}\\
&\propto &\exp \left\{ { - \frac{1}{2}\left[ {{\alpha ^2}\left( {{\bf u}'{\Lambda ^{ - \frac{1}{2}}}\Sigma _y^{ - 1}{\Lambda ^{- \frac{1}{2}}}{\bf u}+\frac{1}{{{c_1}}}} \right)} \right.} \right.\\
&&\hspace{0.73cm}\left. {\left. { - 2\alpha {\bf u}'{\Lambda ^{ - \frac{1}{2}}}\Sigma _y^{ - 1}Z} \right]} \right\}.
\end{eqnarray*}
Thus, $ {\alpha}\left|{\bf y},{\bf u},{\blambda},{\eta_{-\alpha}}\right.\sim{\rm N}\left({\mu_{\alpha}^*},{h_{\alpha}^{-1}}\right) $ where,
\begin{eqnarray*}
    &&{\mu_{\alpha}^*}={h_{\alpha}^{-1}}{{\bf u}^\prime}{\Lambda^{-\frac{1}{2}}}{\Sigma_{y}^{-1}}Z,\\
    &&{h_{\alpha}}={{\bf u}^\prime}{\Lambda^{-\frac{1}{2}}}{\Sigma_{y}^{-1}}{\Lambda^{-\frac{1}{2}}}
    {\bf u}+\frac{1}{c_1}.
\end{eqnarray*}
%============================================
\item[$\bullet$] \textbf{Full conditional distribution of parameter $ {\sigma^2} $}\\
By assume that  ${\bf b}={\bf y}-X\bbeta-\alpha{\Lambda^{-\frac{1}{2}}}{\bf u}=Z-\alpha{\Lambda^{-\frac{1}{2}}}{\bf u}$ and ${\bf B}={\Lambda^
{-\frac{1}{2}}}{C_{\theta_w}}{\Lambda^{-\frac{1}{2}}}+{\omega^2}{{\rm I}_n}$ we have
\begin{eqnarray*}
\pi\left( {{\sigma ^2}\left| {{y_I},{\bf u},\blambda ,{\eta _{ - {\sigma ^2}}},\mathscr{D}} \right.} \right) &\propto &\pi \left( {{\sigma ^2}\left| {{\bf y},{\bf u},\blambda ,{\eta _{ - {\sigma ^2}}}} \right.} \right)\\
&\propto &p\left( {{\bf y}\left| {{\bf u},\blambda ,\eta } \right.} \right)\pi \left({\sigma ^2} \right)\\
&\propto &\exp\left\{-\frac{1}{2{\sigma^2}}{{\bf b}^\prime}{{\bf B}^{-1}}
    {\bf b}\right\}{\left|{\sigma^2}{\bf B}\right|^{-\frac{1}{2}}}\\
&\times & {\left(\frac{1}{\sigma^2}\right)^{{c_2}-1}}\exp\left\{-\frac{c_3}{\sigma^2}\right\}.
\end{eqnarray*}
Since this full conditional distribution does not have any standard form, we use the Metropolis-Hastings method.
%============================================
\item[$\bullet$] \textbf{Full conditional distribution of parameter $ {\omega^2} $}\\
By recalling $\bf b$ and $\bf B$ from the above item, we also have
\begin{eqnarray*}
\pi \left( {{\omega ^2}\left| {{y_I},{\bf u},\blambda ,{\eta _{ - {\omega ^2}}},\mathscr{D}} \right.} \right) &\propto &\pi \left( {{\omega ^2}\left| {{\bf y},{\bf u},\lambda ,{\eta _{ - {\omega ^2}}}} \right.} \right)\\
&\propto &p\left( {{\bf y}\left| {{\bf u},\lambda ,\eta } \right.} \right)\pi \left( {{\omega ^2}} \right)\\
&\propto &\exp\left\{-\frac{1}{2{\sigma^2}}{{\bf b}^\prime}{{\bf B}^{-1}}
    {\bf b}\right\}{\left|{\sigma^2}{\bf B}\right|^{-\frac{1}{2}}}\\
&\times & {\omega ^{ - 2}}\exp \left\{ { - \frac{1}{2}\left[ {c_4^2{\omega ^{ - 2}} + c_5^2{\omega ^2}} \right]} \right\}
\end{eqnarray*}
and hence $ {\omega^2}\left|{\bf y},{\bf u},\blambda,{\eta_{-\omega^2}}\right.\sim{\rm GIG}\left(-\frac{n}{2},\sqrt{{c_{4}^2}+\frac{{{\bf b}^\prime}{{\bf B}^{-1}}{{\bf b}}}{\sigma^2}},{c_5}\right) $.
%============================================
\item[$\bullet$] \textbf{Full conditional distribution of parameter $ \nu $}
\begin{eqnarray*}
\pi \left( {\nu \left| {{y_I},{\bf u},\blambda ,{\eta _{ - \nu }},\mathscr{D}} \right.} \right) &\propto &\pi \left( {\nu \left| {{\bf y},{\bf u},\blambda ,{\eta _{ - \nu }}} \right.} \right)\\
&\propto &p\left( {{\bf y}\left| {{\bf u},\blambda ,\eta } \right.} \right)\pi \left( {\blambda \left| {{\bf u},\eta } \right.} \right)\pi \left( \nu  \right)\\
&\propto &\pi \left( {\blambda \left| \eta  \right.} \right)\pi \left( \nu  \right)\\
&\propto &\exp \left\{ { - \frac{1}{2}{{\left( {\bpsi  + \frac{\nu }{2}J} \right)}^\prime }{{\left( {\nu {C_{{\theta _\lambda }}}} \right)}^{ - 1}}\left( {\bpsi  + \frac{\nu }{2}J} \right)} \right\}\\
&\times &\frac{1}{{{\nu ^{1 + \frac{n}{2}}}}}\exp \left\{ { - \frac{1}{2}\left[ {c_6^2{\nu ^{ - 1}} + c_7^2\nu } \right]} \right\}.
\end{eqnarray*}
In this case, the Metropolis-Hastings method can be used for sampling from this full conditional as well. Recall that this full conditional also can be represent in terms of ${\bf a}=\ln\blambda+\frac{\nu}{2}{\rm J}=\bpsi +\frac{\nu}{2}{\rm J}$ in the penultimate line which is used in the next item. Therefore,
\begin{eqnarray*}
\pi \left( {\nu \left| {{y_I},{\bf u},\blambda ,{\eta _{ - \nu }},\mathscr{D}} \right.} \right) \propto 
\frac{1}{{{\nu ^{1 + \frac{n}{2}}}}}\exp \left\{ { - \frac{1}{2}\left[ {c_6^2{\nu ^{ - 1}} + c_7^2\nu  + {{\bf{a}}^\prime }{{\left( {\nu {C_{{\theta _\lambda }}}} \right)}^{ - 1}}{\bf{a}}} \right]} \right\}.
\end{eqnarray*}
%============================================
\item[$\bullet$] \textbf{Full conditional distributions of the correlation parameters 
  ${\theta_\lambda}$ and $\theta_w$}\\
In terms of $\bf a$, $\bf b$ and $\bf B$, we have
\begin{eqnarray*}
&&\hspace{-0.7cm}
    p\left({\theta_w}\left|{\bf y},{\bf u},\blambda,
    {\eta_{-{\theta_w}}}\right.\right)\propto
    \exp\left\{-\frac{c_8}{{med}\left(d\right)}{{\theta_w}}\right\}\\
    &&\hspace{2cm}
    \times\exp\left\{-\frac{1}{2{\sigma^2}}{{\bf b}^\prime}{{\bf B}^{-1}}{\bf b}\right\}
    {\left|{\sigma^2}{\bf B}\right|^{-\frac{1}{2}}}\\
    &&\hspace{2cm}
    \times\exp\left\{-\frac{1}{2}{{\bf u}^\prime}{C_{\theta_w}^{-1}}{\bf u}\right\}
\end{eqnarray*}
%and since
%\begin{eqnarray*}
%    {\rm Var}\left({Y}\left|u,\blambda,\eta\right.\right)=
%    {\sigma^2}\left({\Lambda^{-\frac{1}{2}}}{C_{\theta_w}}{\Lambda_{I}^{-\frac{1}{2}}}
%    +{\omega^2}{{\rm I}_n}\right)={\sigma^2}B
%\end{eqnarray*}
and
\begin{eqnarray*}
    &&\hspace{-0.7cm}
    p\left({\theta_\lambda}\left|{\bf y},{\bf u},\blambda,
    {\eta_{-{\theta_\lambda}}}\right.\right)\propto
    \exp\left\{-\frac{c_9}{{med}\left(d\right)}{\theta_\lambda}\right\}\\
    &&\hspace{2cm}
    \times\exp\left\{-\frac{1}{2}{{\bf a}^\prime}
    {\left(\nu{C_{\theta_\lambda}}\right)^{-1}}{\bf a}\right\}.
\end{eqnarray*}
These full conditional distributions are of nonstandard form, so a Metropolis-Hastings
step would be used. 
\end{itemize}

\newpage
\section*{Appendix B: Rainfall Data}
\label{AppB}
\begin{table}[h]
\centering
\caption{\footnotesize Rainfall amounts of Fars province measured in inch, geographical locations, and elevation in meter.}
\begin{tabular}{l c c c c}
\toprule
station &precipitation&longitude &latitude&elevation\\
\cmidrule(lr){1-5}
Abadeh &1.09345684&52.40&31.11&2030 \\
Arsanjan &0.48489401&53.16&29.56&1703\\
Bavanat &0.21459461&53.40&30.28&2231 \\
Darab &0&54.17&28.47&1098 \\
Eqlid &1.29816493&52.38&30.54&2300\\
Estahban &0.41391564&54.02&29.05&1609 \\
Farashband & 0.37737203 &52.06&28.48&782 \\
Fasa &0.47246634 &53.41&28.56&1288 \\
Firuzabad &0&52.33&28.53&1362 \\
Gerash & 0.29725655 &54.15&27.69&403\\
Jahrom &0.21737661&53.32&28.29&1082\\
Kavar &0.27805494&52.65&29.16&651 \\
Kazerun &0.31700190 &51.39&29.36 &860\\
Kherameh &0.26663226&53.29&29.63&875\\
Khonj &0.55668826 &53.40&27.98&511\\
Khorrambid &0.77678033&53.09&30.35&2251\\
Lamerd &0&53.12&27.22&405\\
Larestan &0&54.17&27.42&792\\
Mamasani & 0.83764997 &51.32&30.04&972\\
Marvdasht & 0.51173630 &52.54&29.56&1605\\
Mohr & 0.23188202&52.88&27.55 &659\\
Neyriz &0.58328311 &54.20&29.12&1632\\
Pasargad &0.74427868 &53.21&30.19&1614\\
Qir-o-Karzin &0&53.03&28.28&746\\
Rostam &1.49020491&51.51&30.25& 864\\
Sarvestan &0.42793923&53.21&29.27&719\\
Sepidan &1.61606073&52.00&30.14&2201\\
Shiraz&0.68990878 &52.36&29.32&1484\\
Zarghan&0.22756008 &52.43&29.47&1596\\
Zarrindasht&0.34310578  &54.25&28.21&1029\\
\bottomrule
\end{tabular}
\label{66tab}
\end{table}

\newpage
\section*{Appendix C: Details On the Used Distributions}
An $n$-dimensional random vector $Y$ is said to have a multivariate closed skew-normal distribution, denoted by $CSN_{n,m}\left(\bmu,\Sigma,D,{\bf v},\Theta \right)$, if its density is
$\phi_n\left({\bf y};\bmu,\Sigma \right) \Phi_m\left(D\left( {\bf y}-\bmu\right) ;{\bf v},\Theta \right) / \Phi_m\left({\bf 0};{\bf v},\Theta +D\Sigma'D \right)$, where $\bmu \in {\bR}^n$, ${\bf v} \in {\bR}^m$ and $\Sigma \in {\bR}^{n\times n}$ and $\Theta \in {\bR}^{m\times m}$ are both covariance matrices, $D \in {\bR}^{m\times n}$, $\phi_n\left({\bf y};\bmu,\Sigma \right)$ and $\Phi_n\left({\bf y};\bmu,\Sigma \right)$ are the probability density function (pdf) and cumulative distribution function (cdf), respectively, of the $n$-dimensional normal distribution with mean vector $\bmu$ and covariance matrix $\Sigma$.

Arellano-Valle and Azzalini (2006) presented a family of skew-normal distributions which unifies a plethora of $SN$ distributions including $CSN$ with no over-parametrization problem. According to their approach, an $n$-dimensional random vector $Y$ has a multivariate unified skew-normal distribution, denoted by $SUN_{n,m}\left(\bmu, \Sigma, \Gamma,{\bf v}, \Delta \right)$, if its density is
\begin{eqnarray*}
\phi_n\left({\bf y};\bmu,\Sigma \right)\Phi_m\left(\Gamma'{\Sigma^{-1}}\left( {\bf y}-\bmu\right) ;{\bf v},\Delta-\Gamma' {\Sigma^{-1}}\Gamma\right) / \Phi_m\left({\bf 0};{\bf v},\Delta \right),
\end{eqnarray*}
where $\Delta \in {\bR}^{m\times m}$ is a correlation matrix, $\Gamma \in {\bR}^{n\times m}$. The $SUN$ and $CSN$ classes are equivalent when $\Delta=\Theta+D\Sigma D'$  and $\Gamma=\Sigma D$.

Let $\left\{ {G\left({\bf s} \right),s \in R \subseteq {\bR^d}} \right\}$, $d \ge 1$, be
a spatial, ergodic, stationary, zero-mean Gaussian process with stationary covariance function $c\left({\bf h} \right)$ and denoted the covariance matrix of the random vector ${\bf G} = {\left( {G\left( {{s_1}} \right), \ldots ,G\left( {{s_n}} \right)} \right)^\prime }$ by $\Sigma$. Also consider $V\sim {N_n}\left({\bf 0}, {C_\theta}\right)$ and $U\sim {TN_{n}}\left({\bf 0}; {\bf 0}, {C_\theta}\right)$  are independent in which, ${TN_{n}}\left( {\bf c}; \bmu, \Sigma\right)$ denotes the $N_n({\bmu} ,\Sigma)$ distribution truncated below at a point ${\bf c}$. Then if ${{\bf W}_\lambda }\mathop  = \limits^d \alpha {\Lambda ^{ - \frac{1}{2}}}{\bf U} + \sigma {\Lambda ^{ - \frac{1}{2}}}{\bf V}$ and $\bf T$ be a normal vector of dimension $n$, we have 
\begin{eqnarray*}
\left( \begin{array}{l}
{\bf T}\\
{\bf G}
\end{array} \right)\left| {\blambda  \sim {N_{2n}}\left( {{\bf 0},\left[ {\begin{array}{*{20}{c}}
{{C_\theta }}&{\alpha {C_\theta }{\Lambda ^{ - \frac{1}{2}}}}\\
{\alpha {\Lambda ^{ - \frac{1}{2}}}{C_\theta }}&{{\Sigma _w}}
\end{array}} \right]} \right)} \right..
\end{eqnarray*}
Now, we define a $SUGLG$ process $\{W_\lambda(s)\}$ based on the following hierarchical representation:
\begin{enumerate}
\item $\left[ {{W_\lambda }\left( {\bf s} \right)\left| {\lambda \left({\bf s} \right)} \right.} \right]\mathop  = \limits^d \left[ {G\left( {\bf s} \right)\left| {{\bf T} >{\bf 0},\lambda \left( {\bf s} \right)} \right.} \right]$,
\item ${\lambda \left({\bf s} \right)}$ is a log-Gaussian stochastic process.
\end{enumerate}

%\section*{Acknowledgment}
%This is the acknowledgment section of the paper. It is unnumbered.

\end{document}